\documentclass{aa}
\usepackage{psfig}
\usepackage{graphicx}
\usepackage{natbib}
\title{Infrared spectroscopy of solid CO-CO$_2$ mixtures and layers}
\subtitle{}
\author{F.A. van Broekhuizen\inst{1}, I.M.N. Groot\inst{1},
H.J. Fraser\inst{1,*}, E.F. van Dishoeck\inst{1,2}, and S.Schlemmer\inst{1}}
\offprints{E.F. van Dishoeck\\e-mail: ewine$@$strw.leidenuniv.nl\\
$^*$ Current adress: Department of Physics, University of Strathclyde, John Anderson Building 107 Rottenrow, Glasgow G4 0NG, Scotland \\$^a$http://www.strw.leidenuniv.nl/\~{}lab/databases/
}
\institute{$^1$ Raymond and Beverly Sackler Laboratory for Astrophysics, Leiden
Observatory, P.O. Box 9513, 2300 RA Leiden, The Netherlands\\
$^2$ Leiden Observatory, P.O. Box 9513, 2300 RA Leiden, The Netherlands}
\date{November 30 2005, accepted in A\&A}
\authorrunning{F.A. van Broekhuizen et al.}
\titlerunning{Infrared spectroscopy of.....}
\begin{document}
\abstract{The spectra of pure, mixed and layered CO and CO$_2$ ices
have been studied systematically under laboratory conditions using
Transmission-Absorption Fourier Transform infrared spectroscopy. This
work provides improved resolution spectra (0.5\,cm$^{-1}$) of the
CO$_2$ bending and asymmetric stretching mode, as well as the CO
stretching mode, extending the existing Leiden database$^a$ of
laboratory spectra to match the spectral resolution reached by modern
telescopes and to support the interpretation of the most recent data
from the Spitzer Space Telescope. It is shown that mixed and layered
CO and CO$_2$ ices exhibit very different spectral characteristics,
which depend critically on thermal annealing and can be used to
distinguish between mixed, layered and thermally annealed CO-CO$_2$
ices. CO only affects the CO$_2$ bending mode spectra in mixed ices
below 50\,K under the current experimental conditions, where it
exhibits a single asymmetric band profile in intimate mixtures. In all
other ice morphologies the CO$_2$ bending mode shows a double peaked
profile, similar to that observed for pure solid CO$_2$. Conversely,
CO$_2$ induces a blue-shift in the peak-position of the CO stretching
vibration, to a maximum of 2142\,cm$^{-1}$ in mixed ices, and
2140--2146\,cm$^{-1}$ in layered ices. As such, the CO$_2$ bending
mode puts clear constraints on the ice morphology below 50\,K, whereas
beyond this temperature the CO$_2$ stretching vibration can
distinguish between initially mixed and layered ices. This is
illustrated for the low-mass young stellar object HH\,46, where the
laboratory spectra are used to analyse the observed CO and CO$_2$ band
profiles and try to constrain the formation scenarios of CO$_2$.
 }
\maketitle
\section{Introduction}
\label{introduction}
CO$_2$ is one of the most abundant components of interstellar ice
after H$_2$O and CO \citep{Gibb2004,Boogert2004,Nummelin2001,Gerakines1999,Whittet1998}
and has been observed in the solid-state on lines-of-sight towards
a variety of high- and low-mass stars, field stars and galactic
centre sources. Surprisingly little variation in abundance exists
between  these objects, generally ranging from 10 to 23\,\% with respect
to H$_2$O-ice, although values of 34 to 37\,\% have been reached
for some low- and intermediate mass sources \citep{Nummelin2001,Boogert2004,Pontoppidan2005}. In the gas phase, observations imply that the CO$_2$ abundance is a factor of $\sim$\,100 less than in the solid state \citep{vanDishoeck1996,Boonman2003}, indicating that CO$_2$ forms in the
solid-phase. The most popular formation mechanism is via energetically mediated reactions such as UV photolysis of mixed H$_2$O-CO ices \citep{Watanabe2002}. However, the
fact that CO$_2$ ice is also observed towards the quiescent clouds
in front of Elias\,16, where there is little if any UV- or cosmic ray-induced chemistry, suggests that no energetic ice-processing is
required for CO$_2$ formation, so surface reactions involving
CO must play a key role \citep{Whittet1998,Roser2001}.
Whichever formation route is invoked, a direct link is implied between the
location of CO and CO$_2$ within
interstellar ices.\\
\indent Recent high resolution observations of solid CO towards a
large sample of embedded objects using the VLT-ISAAC spectrometer
show that 60-90\,\% of solid CO in interstellar ices resides in a nearly pure form, and that a significant number of
the sources show evidence for CO in a H$_2$O-rich environment, at
a minimum abundance of 19\,\% with respect to the pure CO
component \citep{Pontoppidan2003,Boogert2004,Fraser2004}. Given these constraints, CO$_2$ may form via surface reactions on the pure CO-ice layer, resulting in a bi-layered ice-structure. Alternatively, if
CO$_2$ forms from UV- and cosmic ray-induced reactions involving CO
in the H$_2$O-rich environment, one can envisage that CO$_2$ will
reside in a H$_2$O-rich ice without being in direct contact with CO. The latter situation has been
simulated in detail in the laboratory by, for example, \citet{Gerakines1995} and \citet{Ehrenfreund1999}, who recorded infrared (IR) spectra of
mixed ices. \\
\indent The key aim of this paper is to establish the influence CO may have on the CO$_2$ spectral features and vice versa. In addition to the spectroscopic features of CO at 2139 and 2136 cm$^{-1}$ in pure and H$_2$O-rich ice environments respectively, a third component of the interstellar CO-ice band has been detected at around 2143 cm$^{-1}$ along many
lines of sight. In the VLT-ISAAC and other surveys, this band was attributed
to the TO-component of the pure CO ice feature, which under certain conditions is split into two sharp features \citep{Collings2003LO,Pontoppidan2003}. However, \citet{Boogert2002b} assign this band to CO in a CO$_2$ dominated environment. This paper therefore also analyses the presence or absence of the 2143 cm$^{-1}$ feature in the spectra of CO-CO$_2$ laboratory ices, as a function of the morphology (i.e., the configuration of CO and CO$_2$ in the CO-CO$_2$ ice system: either mixed or separated in two spatially distinct layers) and temperature, to find the origin of this band.
\\
\indent Solid CO$_2$ can be observed in interstellar spectra by its 4.3\,$\mu$m asymmetric stretching ($\nu_3$) and 15\,$\mu$m bending ($\nu_2$) modes. The $\nu_2$ mode is known to
be very sensitive to the local ice environment but is weaker than the
stretching mode. The higher sensitivity of the recently launched
Spitzer Space Telescope, compared to that of its predecessor the Infrared
Space Observatory (ISO), has made it possible to conduct a detailed analysis
of the $\nu_2$(CO$_2$) band over a wider range of
interstellar objects. First results show that the CO$_2$ abundances are similar to or higher than those deduced with ISO \citep{Watson2004,Boogert2004,Pontoppidan2005}. Thus, there is the possibility to combine the new observational data of solid CO$_2$ with VLT observations of CO to place constraints on the CO$_2$ ice environment. \\
\indent Previous laboratory studies have dealt with mixtures of binary ices composed of CO
and CO$_2$ \citep{Ehrenfreund1996,Ehrenfreund1997,Elsila1997}. These proved very useful for interpreting interstellar ice observations. However,
more systematic studies including layered ices are necessary to link the spectral characteristics to
the interstellar environments mentioned above. This work therefore presents the effects of CO on the spectroscopy
of the CO$_2$-bending and stretching modes as well as the influence
of thermal annealing in pure, mixed and layered ice systems. The
data are taken at an increased spectral resolution of 0.5\,cm$^{-1}$ with respect to the previous laboratory studies, in order to resolve the CO$_2$-bending mode and the CO-stretching vibration and meet the accuracy of the 
VLT-ISAAC and Spitzer
data as well as future mid-infrared spectra to be obtained with the James Webb Space Telescope. The experimental details are explained in Sect.~\ref{exp}.
The spectroscopic behaviour of CO and CO$_2$ are presented and
discussed in Sects.~\ref{results} and~\ref{discussion},
respectively, as a function of the temperature and ice composition. In Sect.~\ref{astroimpl} the astrophysical implications
are discussed and one example of a comparison between the new
experimental data with observations is given.
\section{Experimental Procedure}
\label{exp} 
All experiments were conducted in a high vacuum (HV)
chamber described in detail elsewhere \citep{Gerakines1995}, at a base-pressure below 2$\times$10$^{-7}$ Torr. Ices of $^{12}$C$^{16}$O
(Praxair 99.997\% purity) and $^{12}$C$^{16}$O$_2$ (Praxair
99.997\% purity) were grown on the surface of a CsI
window, pre-cooled to 15\,K, via effusive dosing at a growth rate of
$\sim$\,10$^{16}$\,molec\,cm$^{-2}$s$^{-1}$, directed at 45
degrees to the surface normal. Transmission Fourier Transform
Infrared spectra of the ice systems were recorded at 0.5\,cm$^{-1}$ resolution at fixed temperatures between 15 to 100\,K, using a total of 128 scans between 4000--400\,cm$^{-1}$ and a zero filling factor of 4. Each recording lasted 20 min, and was started directly after the temperature had equilibrated, which took 2\,min.\\
\indent The pure and layered ice structures were grown in situ, via single and sequential dosing from CO and CO$_2$ gas bulbs,  filled to a total pressure of 10\,mbar, prepared
on a separate glass vacuum manifold, with a base-pressure of $\sim$\,10$^{-5}$
mbar. Mixed ices were prepared by dosing gas from pre-mixed CO and
CO$_2$ bulbs. \\
\indent The Full Width at Half Maximum (FWHM) and peak-position of
the CO-stretching, CO$_2$-bending and CO$_2$ asymmetric stretching
mode were determined using the Levenberg-Marquardt non-linear
least square fitting routine within Origin 7.0. Uncertainties in the peak-centre position and FWHM were typically less than 0.05\,cm$^{-1}$, except close to the desorption temperature where uncertainties may be larger.  \\
\indent Previous work on the spectroscopy of CO$_2$-ice has indicated that the ice structure (i.e. amorphous or crystalline) influences its spectral profile \citep{Falck1986}, although some uncertainty exists as to the extent of the crystallinity of low temperature vapour deposited CO$_2$-ice \citep{Falck1986,Sandford1988,Sandford1990}. Consequently, optical constants have not been derived for the ices studied here. Before doing so, a systematic study is required of the degree of crystallinity in pure, vapor-deposited CO$_2$ ices and the influence of these phases on the CO$_2$ spectroscopy, which will be the topic of future work.  \\
\indent The range of ices studied here are summarised in Table~\ref{icecomp}. The relative concentrations of CO and CO$_2$ range from 0.1 to 2 CO/CO$_2$, relevant for comparison with observations of high- and low-mass YSO's, which show CO/CO$_2$ column density ratios of 0.1--1.3 \citep{Gibb2004}. All of the raw laboratory spectra discussed here have been made available through the Internet at http://www.strw.leidenuniv.nl/\~{}lab/mixed\_{}layered\_{}CO\_{}CO2/\\
\begin{table*}
\caption{Ice compositions}
\centering
\begin{tabular}{l|l|c|c|l|l}
\hline
Ice  & Nomenclature & \multicolumn{2}{|c|}{Conc. ratio} & Exposure & Approx. ice \,$^{a}$\\
morphologies     &       &  CO  &  CO$_2$  & (s)      & thickness (L) \\
\hline
pure & CO & 1 & - & 60 & 600  \\
     & CO$_2$ & - & 1 & 60 & 600  \\
\hline
layered & 1/1 CO/CO$_2$  & 1 & 1  & 60 + 60  & 1200 \\
        & 1/1 CO$_2$/CO  & 1 & 1  & 60 + 60  & 1200  \\
        & 1/2 CO/CO$_2$  & 1 & 2  & 60 + 120 & 1800 \\
        & 2/1 CO$_2$/CO  & 2 & 1  & 120 + 60 & 1800 \\
        & 3/1 CO$_2$/CO  & 3 & 1  & 180 + 60 & 2400 \\
        & 10/1 CO$_2$/CO & 10 & 1 & 600 + 60 & 6600 \\
\hline
mixed   & 1:1 CO:CO$_2$  & 1 & 1  & 120 & 1200 \\
        & 2:1 CO:CO$_2$  & 2 & 1  & 180 & 1800  \\
        & 1:10 CO:CO$_2$ & 1 & 10 & 660 & 6600  \\
\hline
\end{tabular}
\begin{itemize}
\item[]\footnotesize{The nomenclature adopted uses {\it A}:{\it B} to denote mixtures and {\it
A}/{\it B}\, to indicate layered structures with {\it A} on top of {\it B}. $^{a}$ 1\,L = 10$^{15}$\,molec\,cm$^{-2}$\,$\approx$\,1 mono-layer.}
\end{itemize}
\label{icecomp}
\end{table*}
\section{Results}
\label{results}
%
\begin{figure*}
\centering
\includegraphics[width=18cm]{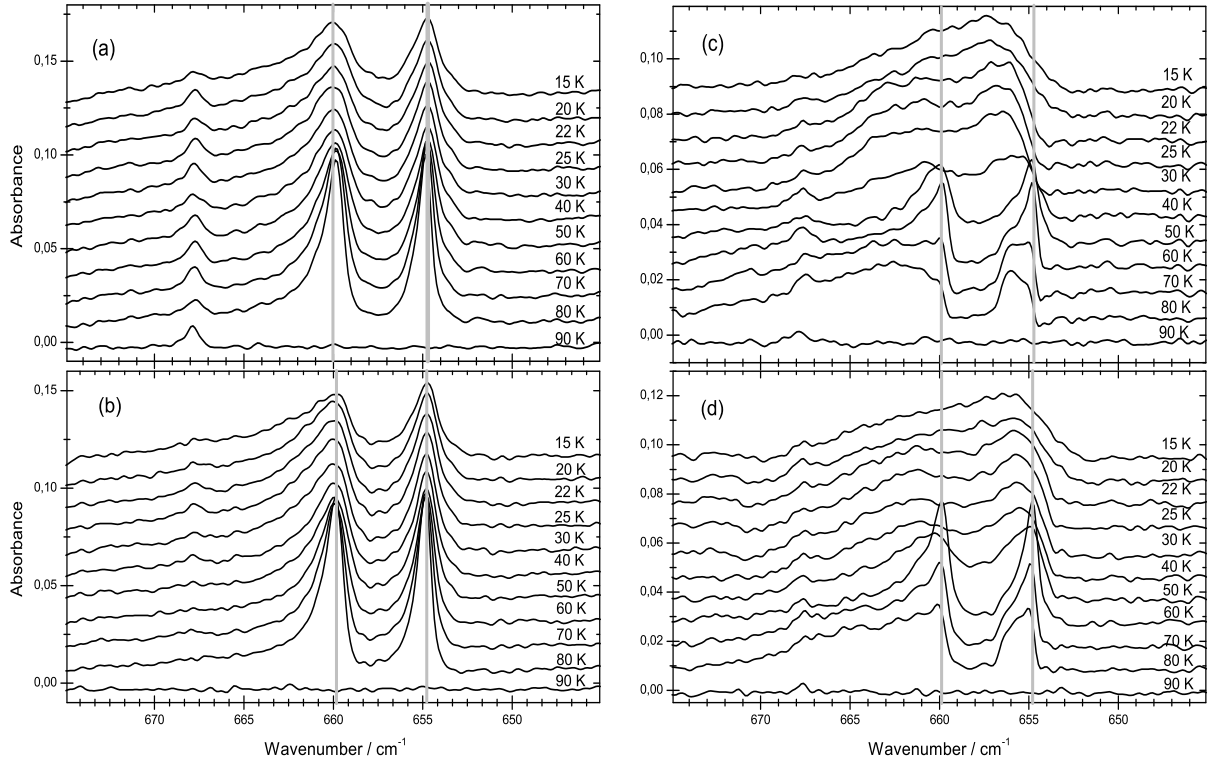}
\caption{The thermal evolution of the CO$_2$-bending mode for ({\bf a}) pure CO$_2$-ice, ({\bf b}) a 1/1 CO/CO$_2$ layered ice, and ({\bf c}) 2:1 CO:CO$_2$ and ({\bf d}) 1:1 CO:CO$_2$ ice mixtures, from 15--90\,K. The two vertical grey lines mark the positions of the two peaks of the bending mode at 15\,K in pure CO$_2$-ice.}
\label{mml_co2b}
\end{figure*}
\subsection{CO$_2$-bending mode}
\label{co2b}
Fig.~\ref{mml_co2b} shows the spectra of $\nu_2$(CO$_2$) at 15\,K
in pure and some example mixed and layered ices (with CO), together
with the thermal evolution of each band. In the pure ice
(Fig.~\ref{mml_co2b}a), the $\nu_2$ spectrum at 15\,K has two
components, peaking at 654.7 and 659.8 cm$^{-1}$ (marked by grey
lines) with FWHM of 1.8 and 3.1 cm$^{-1}$, respectively, showing a
Davidov splitting due to a highly ordered ice structure
\citep[e.g.,][]{Molvib,Sandford1990}. This profile changes gradually
beyond 40\,K as both peak intensities increase while their FWHM
decrease until the point at which CO$_2$ desorbs, in agreement with
previous findings of \citet{Sandford1990}.  The
integrated area is observed to increase with temperature, but by no
more than 20\%. This increase is consistently found in the bending mode
spectra during the warm-up of pure and layered ices. This is in
contrast with the spectral behaviour of the CO$_2$ stretching mode (see
Sect.\ref{co2s}), which does not show any increase in integrated band
strength but stays constant within 5\%. The spectral
profile and thermal behaviour is very similar in 1/1 CO/CO$_2$, shown
in Fig.~\ref{mml_co2b}b. Interestingly, this appears to be independent
of whether CO$_2$ is deposited on top or underneath the CO-ice layer
(data not shown).\\ 

\indent Conversely, when CO$_2$ is mixed with CO,
the $\nu_2$ spectrum changes drastically \citep[see
also][]{Ehrenfreund1997,Elsila1997}. At 15\,K, the 2:1 and 1:1
CO:CO$_2$ ices, Figs.~\ref{mml_co2b}c and \ref{mml_co2b}d
respectively, show one broad asymmetric band without any
sub-structure, peaking at between 655 and 657\,cm$^{-1}$. However,
from the present experiments it is clear that at 22\,K, some
sub-structure becomes visible, which evolves to a doublet beyond
40\,K. This doublet is similar to that of pure CO$_2$-ice, but the
peaks remain slightly broader and more asymmetric.  Under the present
experimental conditions, CO$_2$ desorbs between 80 and 90\,K in all
the ices studied. The exception is 1:10 CO:CO$_2$. Its 15\,K spectrum
exhibits two peaks centred at 654.7 and 659.8 cm$^{-1}$ (FWHM of 3.6
and 6.1\,cm$^{-1}$) and its thermal behaviour is similar to that of
pure CO$_2$-ice. However, its desorption is retarded to 90--100\,K. A
small artifact is occasionally evident in Fig.~\ref{mml_co2b} at
667.8\,cm$^{-1}$, which is clearly not associated with CO or CO$_2$
since it remains in the spectra after CO$_2$ desorbs.  
\subsection{CO$_2$ asymmetric stretching vibration}
\label{co2s}
\begin{figure*}
\centering
\includegraphics[width=18cm]{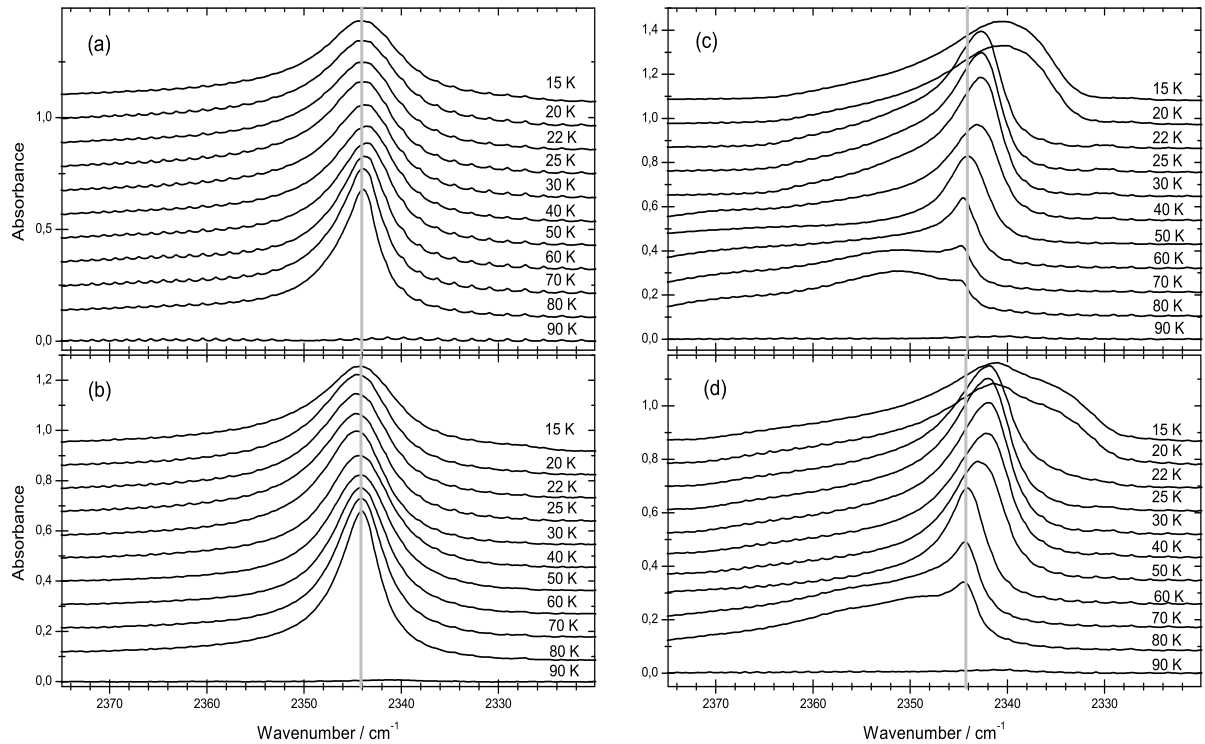}
\caption{The thermal evolution of the asymmetric stretching mode of CO$_2$ in ({\bf a}) pure, ({\bf b}) 1/1 CO/CO$_2$, ({\bf c}) 2:1 CO:CO$_2$ and ({\bf d}) 1:1 CO:CO$_2$, from 15--90\,K. The grey line marks the peak-centre position at 15\,K in pure CO$_2$-ice.}
\label{mml_co2s}
\end{figure*}
Fig.~\ref{mml_co2s} shows the thermal evolution of the $\nu_3$(CO$_2$) for the same four ices as Fig.~\ref{mml_co2b}. As with the $\nu_2$ mode, the trend observed is that almost no spectral differences are seen between the pure and layered ices (Fig.~\ref{mml_co2s}a and b), whereas peak-position and FWHM differences are observed between the pure and the mixed ices (Figs.~\ref{mml_co2s}a and \ref{mml_co2s}c--d). This indicates that it will be difficult to distinguish between pure and layered ices in interstellar spectra. However, since the thermal evolution of mixed and layered ices are different, the comparison between these two cases can help to unravel the environmental history of interstellar ices.\\
\indent In pure CO$_2$-ices at 15\,K, the $\nu_3$(CO$_2$) band peaks at 2344.0\,cm$^{-1}$ (FWHM of 10.6\,cm$^{-1}$), independent of temperature. Its FWHM, however, starts to decrease gradually beyond 40\,K as the peak intensity increases until CO$_2$ desorbs. In mixtures (Figs.~\ref{mml_co2s}c and \ref{mml_co2s}d, respectively) at least three different spectral components are visible due to the higher resolution used compared to previous studies \citep[e.g.,][]{Sandford1990,Ehrenfreund1997}. At 15\,K, the spectral profile is dominated by a broad asymmetric band, peaking at 2339.9\,cm$^{-1}$ (FWHM of 13.3\,cm$^{-1}$), which exhibits a shoulder at around 2334\,cm$^{-1}$ in the case of 1:1 CO:CO$_2$. By 22\,K, the spectrum has evolved into a pure ice like feature, peaking at 2342.5\,cm$^{-1}$, which beyond 50\,K shows the development of a third component at around 2351\,cm$^{-1}$.
%
\begin{figure*}
\centering
\includegraphics[width=18cm]{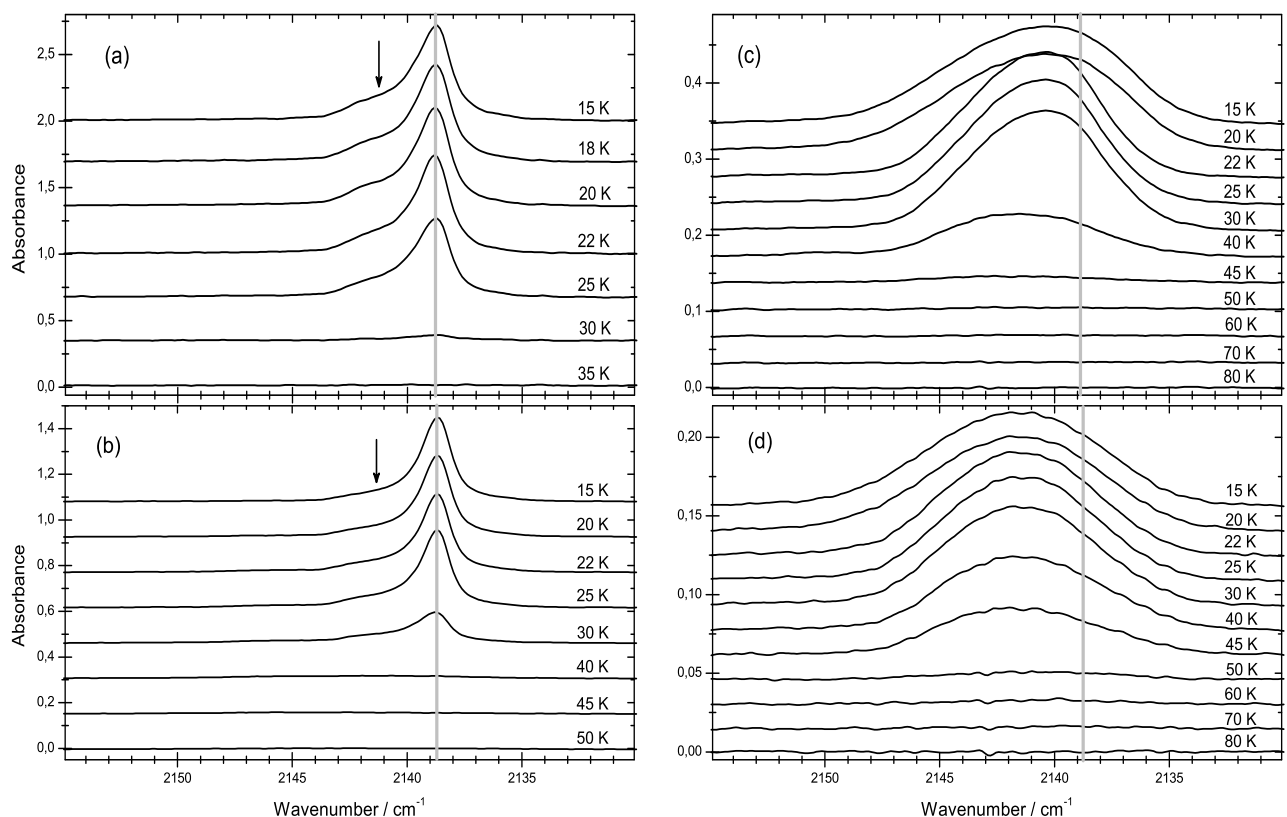}
\caption{The thermal evolution of the CO-stretching vibration in ({\bf a}) pure CO-ice, ({\bf b}) 1/1 CO/CO$_2$, ({\bf c}) 2:1 CO:CO$_2$ and ({\bf d}) 1:1
CO:CO$_2$ from 15--80\,K. The grey line marks the position of the main peak at 15\,K in pure CO-ice. }
\label{mml_cos}
\end{figure*}
%
\subsection{CO-stretching vibration}
\label{co} 
Fig.~\ref{mml_cos} shows the thermal behaviour of the
CO-stretching mode for the same four ices as Fig.~\ref{mml_co2b}. The
spectrum of pure CO-ice (Fig.~\ref{mml_cos}a) peaks at
2138.7\,cm$^{-1}$ (FWHM of 1.6\,cm$^{-1}$) and is independent of
temperature. This differs from the results of \citet[][]{Sandford1988}
who observed a band narrowing with increasing temperature. A shoulder,
visible at $\sim$2141\,cm$^{-1}$ (black arrow), was observed
previously by \citet{Sandford1988} and is discussed in the context of
CO trapping in ices by Fraser et al.\ (in prep., hereafter FR06a). Under
the present experimental conditions, pure CO desorbs between
25-30\,K.

\begin{figure}
\centering
\includegraphics[width=8cm]{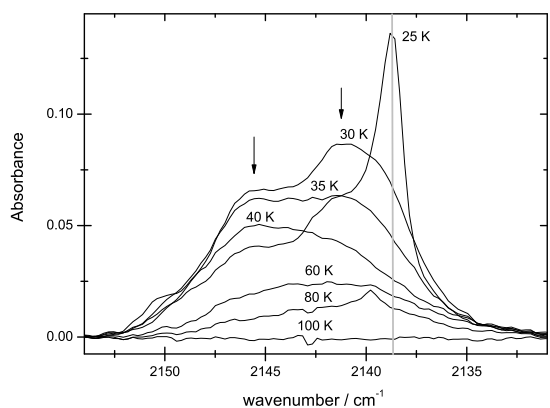}
\caption{The thermal evolution of the CO-stretching vibration in a
10/1 CO$_2$/CO layered ice system, from 25--100\,K.
The thick grey line line marks the peak-centre position
of the main spectral feature in pure CO-ice at 15\,K. The two arrows
indicate the two shoulders that evolve as the temperature increases.}
\label{l_110_cos}
\end{figure}

Again, the CO specta of layered ices are almost
identical to that of pure CO ice. However, beyond 22\,K in 1/1
CO/CO$_2$ (Fig.~\ref{mml_cos}b), the 2141\,cm$^{-1}$ shoulder gets
relatively more pronounced. This effect becomes stronger with
increasing thickness of the CO$_2$ layer and is best seen in
Fig.~\ref{l_110_cos} for the most extreme case of CO$_2$/CO studied
here, 10/1. There, the intensity of the main peak (2138.9\,cm$^{-1}$)
clearly decreases from 22 to 30\,K while the `shoulder' at
2141\,cm$^{-1}$ becomes more intense and a third feature appears at
around 2145.5\,cm$^{-1}$ (both indicated by an arrow). These new
features start decreasing in intensity when CO desorbs from 30\,K
onwards. However, because the 2141\,cm$^{-1}$ feature reduces faster,
this leads to a blue-shift of the overall CO band. From 40--60\,K,
also the 2145.5\,cm$^{-1}$ environment is lost, which shifts the
overall band position back to the red before the remaining CO finally
desorbs (with the CO$_2$) above 80\,K. 

\begin{figure*}
\centering
\includegraphics[width=15cm]{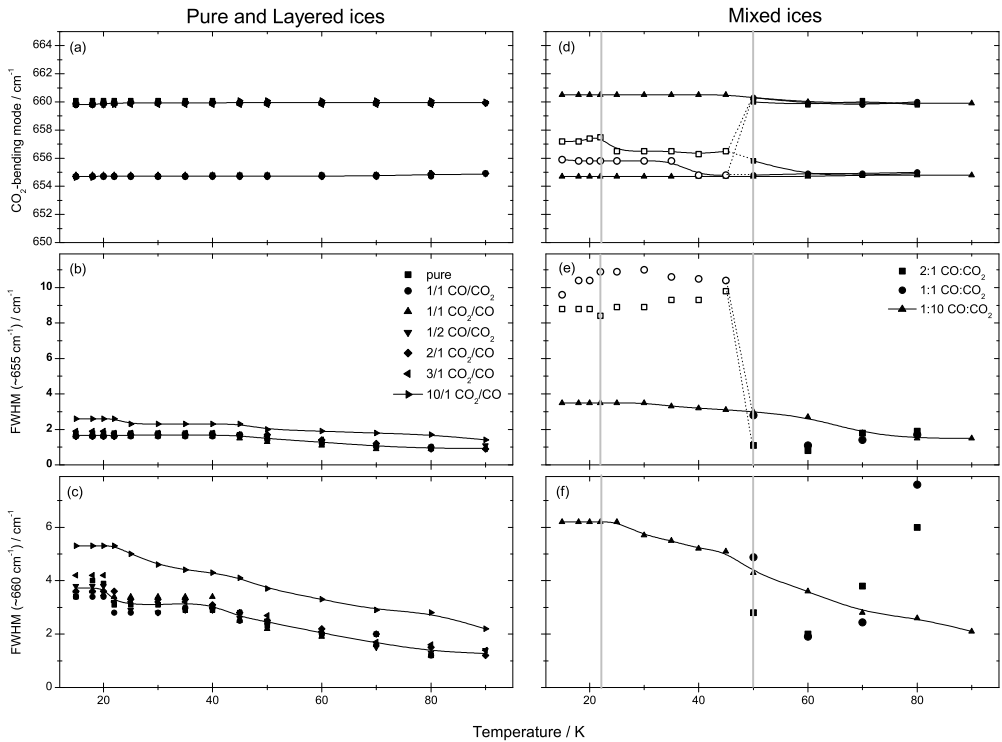}
\caption{The peak-centre positions and FWHM of the main spectral
features of the CO$_2$-bending mode, plotted as a function of ice
temperature. {\bf (a)} Peak-centre positions in the pure and layered
ices with the corresponding FWHM of the feature centred at {\bf (b)}
655\,cm$^{-1}$, and {\bf (c)} 660\,cm$^{-1}$. {\bf (d)} Peak-centre
positions in the mixed ices with the corresponding FWHM of the peak at
{\bf (e)} 655\,cm$^{-1}$, and {\bf (f)} at 660\,cm$^{-1}$. The legend
in {\bf (b)} gives the symbol assignment for all three left-hand
panels, that in {\bf (e)} for all three right-hand panels. Open
symbols indicate the presence of only a single peak, whereas solid
symbols mark the presence of the doublet. In each of the panels,
spline-fits through the data points guide the eye, except in {\bf
(e)} and {\bf (f)} where beyond 50\,K peak positions and FWHM are more
difficult to assign. The two vertical grey lines enclose the
temperature range between 22 and 50\,K (see text for details).}
\label{co2pw}
\end{figure*}

All mixed ices
(Fig.~\ref{mml_cos}c and d) show broadened and blue-shifted spectra
compared to the pure case, centred between 2140.3 and
2141.6\,cm$^{-1}$, with FWHM extending from 5.6 to 9.0\,cm$^{-1}$
\citep[see also][]{Elsila1997}. Equimolar mixtures induce the
strongest blue-shifts. Thermal warming only narrows the band between
20 and 22\,K to a FWHM between 6.5 and 7.6\,cm$^{-1}$ \citep[broader
than predicted by][]{Ehrenfreund1997}. CO desorption starts around
30\,K and continues up to 50--60\,K in 2:1 and 1:1 CO:CO$_2$, or
90--100\,K in the case of 1:10 CO:CO$_2$ (not shown). None of the data
show evidence for an isolated CO-feature centred at 2143\,cm$^{-1}$,
as suggested for CO-CO$_2$ interactions in interstellar ices by
\citet{Boogert2002b}.\\ 
\section{Discussion}
\label{discussion}
\begin{figure*}
\centering
\includegraphics[width=15cm]{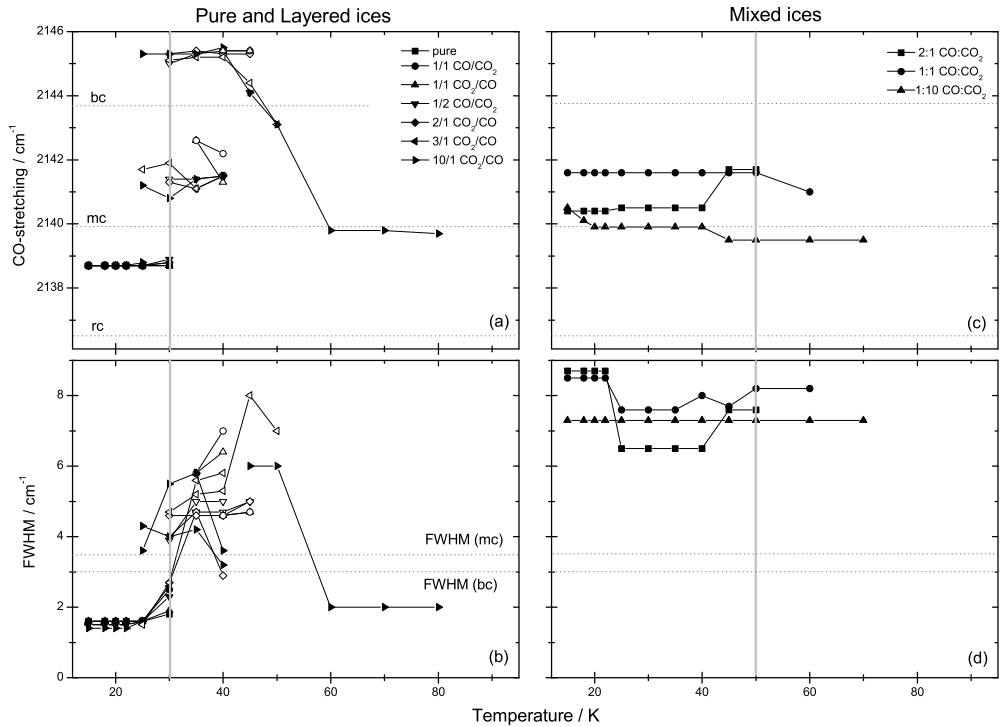}
\caption{The main peak-centre positions and FWHM of the
CO-stretching vibration, plotted as a function of ice
temperature. {\bf (a)} Peak-centre positions in pure and layered ices
with {\bf (b)} their corresponding FWHM. {\bf (c)} Peak-centre
positions in mixed ices with {\bf (d)} their corresponding FWHM. The
legend in {\bf (a)} gives the symbol assignment for the two left-hand
panels, that in {\bf (c)} for the right-hand panels. Solid symbols
indicate strong bands ($S/N>$\,5); open symbols indicate very weak
bands, $S/N$\,1-2 at best. Symbols indicating the same feature are
connected by lines to guide the eye. The vertical grey line in the
left panels marks the temperature at which pure CO desorbs, that in
the right panels marks the temperature at which the single peak
profile of the CO$_2$-bending mode in a mixture converts to a
doublet. The three horizontal (dotted) lines in {\bf (a)} and {\bf
(c)} indicate the line centres of the three components of the
phenomenological fit to the interstellar solid CO-band observed by
Pontoppidan et al. (2003) (see text). The two dotted lines in {\bf
(b)} and {\bf (d)} give the FWHM of the mc and rc of the fitted
components. It should be noted that the FWHM and peak positions of
laboratory spectra can be altered due to grain shape effects.}
\label{copw}
\end{figure*}

The results described in Sect.~\ref{results} provide clear
evidence that the morphology of CO- and CO$_2$-containing ices affects
their spectroscopy. The peak positions and FWHM of all ices studied
are summarised in Figs.~\ref{co2pw} (CO$_2$-bending mode) and
\ref{copw} (CO-stretching mode) to simplify the evaluation of the
spectral changes as functions of temperature and ice morphology. For a
similar evaluation of the CO$_2$-stretching mode, see Appendix A in the
online article. The associated uncertainties of these data have been
calculated (see Sect.~\ref{exp}) but error bars are omitted from the
plots for clarity. The actual ice structure and the physical processes
that give rise to these spectra are discussed by Fraser et al.\ (in
prep., hereafter FR06b). 

In Figs.~\ref{co2pw} and \ref{copw}, the thermally induced spectroscopic changes appear at 22, 30 and 50 K, related to CO bulk-diffusion, the onset of CO desorption
from pure CO ice and the temperature at which pure CO$_2$ ice
undergoes major restructuring, respectively \citep[see for further
discussion][FR06a and FR06b]{Collings2003}. 

It is clear from Figs.~\ref{co2pw}a--c that the $\nu_2$(CO$_2$) spectrum is
always split into two fully resolved peaks where CO$_2$ dominates the
ice matrix, i.e. in pure and layered ices, in (close to) equimolar
mixed ices at $T>50$\,K, and in very dilute CO mixtures (i.e. 1:10
CO:CO$_2$). Despite the very similar band profile of the
$\nu_2$(CO$_2$) in all ices beyond 50 K, the $\nu_3$(CO$_2$) spectra
(see Fig. 2 and Appendix A) indicate that the
line profiles of the vibrational spectra in pure, layered and mixed
CO:CO$_2$ ices are distinct from each other, suggesting that each ice
morphology evolves a distinct structure on annealing. This shows the
importance of observing both the bending and stretching mode
vibrations of CO$_2$, to be able to constrain the interstellar ice
environment of CO$_2$ and its history.\\
\indent Fig.~\ref{copw} shows the main positions and FWHM of the
CO-stretching mode. The temperature-independent band position of CO in
mixtures (Fig.~\ref{copw}c) is consistent with the fact that CO likely
remains bound in the CO$_2$ matrix over the full temperature range,
implying the absence of ice segregation. In layered ices, the fraction
of CO that remains trapped or bound somehow in the CO$_2$ layer
exhibits spectra that are similar to those of mixtures, both in peak
position and in FWHM. This suggest that layered ices are thermally
mixing. This `mixing' apparently does not affect the CO$_2$ ice
structure since the $\nu_2$(CO$_2$) spectra (Figs.~\ref{co2pw}b--c)
show no detectable peak broadening.

The horizontal lines in
Fig.~\ref{copw} indicate the positions and FWHM of the three
phenomenological interstellar CO-ice components
\citep{Pontoppidan2003}. The red component (rc) is associated with a
CO--H$_2$O environment and will not be further addressed here. The
middle component (mc), associated with pure CO-ice, lies close to the
peak-centre position of the CO spectrum in pure and layered laboratory
ices. The blue component (bc) has been invoked as being indicative of
CO:CO$_2$ mixtures in interstellar ice. However, no direct match is
found between the position of this bc and the band positions of the
ices studied here, although both mixed and layered ices do contain a
blue spectral wing. Also, Figs.~\ref{copw}b and \ref{copw}d show that
all ice components between 2141--2146\,cm$^{-1}$ have FWHM 1.5 to 3
times larger than that of the bc. Such evidence does question the
assignment of a 2143\,cm$^{-1}$ feature to CO:CO$_2$ ices but cannot
rule it out completely. Further work will be required on LO-TO
splitting of the CO-stretching mode \citep[a plausible alternative
carrier of the 2143\,cm$^{-1}$ feature, see][]{Collings2003LO} on this
same subset of ice morphologies in the laboratory. A detailed analysis
of high S/N interstellar line profiles is needed of sources that
exhibit a strongly pronounced CO blue wing \citep[e.g., L\,1489,
SVS\,4--9, Reipurth\,50 and RNO\,91,][]{Pontoppidan2003} and a
single-peaked CO$_2$ bending mode profile before the assignment of the
CO 2143\,cm$^{-1}$ band can be fully resolved.  
\section{Astrophysical Implications}
\label{astroimpl}
\begin{figure*}
\centering
\includegraphics[width=15cm]{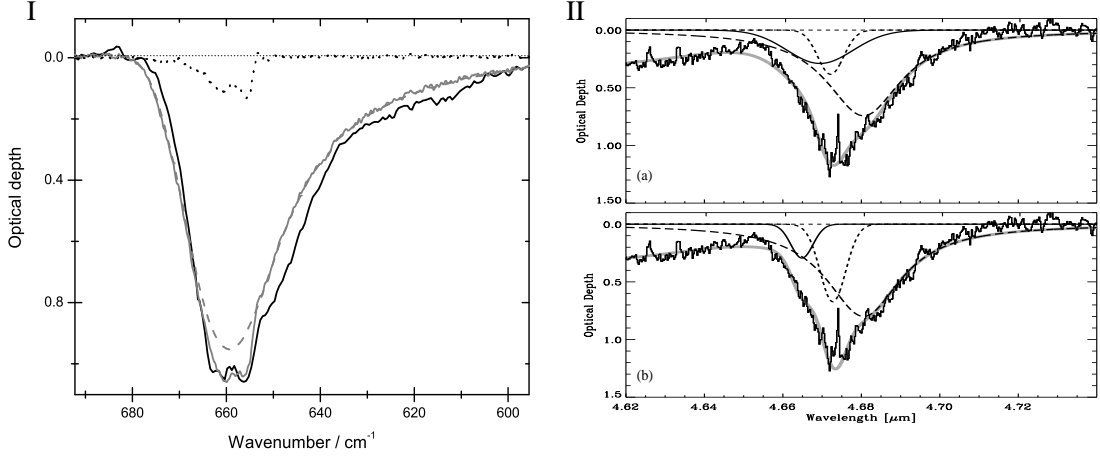}
\caption{{\bf (I, left)} The $\nu_2$(CO$_2$) as observed toward HH\,46 by
Spitzer \citep[black solid line,][]{Boogert2004}, reproduced by a
combination of a laboratory spectrum of 10:1 H$_2$O:CO$_2$ at 10\,K
(light grey dashed line) and a 2:1 CO:CO$_2$ at 40\,K (dotted
line). The resulting match between is shown by the dark grey
line. {\bf (II, right)} The CO-stretching mode as observed towards HH\,46 by
VLT-ISAAC (solid line). Two different fits are shown, adopting
{\bf(IIa)} the laboratory spectrum of 2:1 CO:CO$_2$ at 40\,K (thick
solid line) and of 1:10 CO:H$_2$O at 10\,K (dashed line), and the mc
of solid CO (dotted line), and {\bf(IIb)} the bc, mc, and rc component
of solid CO \citep{Boogert2004}. In both plots, the total fit to the
CO band is shown in grey, including a Gaussian fit to the XCN band to
the blue of the CO stretching mode.}
\label{hh46}
\end{figure*}

\begin{table*}
\caption{Results of the CO-CO$_2$ fit toward HH\,46}
\centering
\begin{tabular}{lcccc}
\hline
Ice component& $N$ (10$^{17}$)\,$^a$&$N$/$N{\rm (CO_2)_{tot}}$  &$N$/$N{\rm (CO)_{tot}}$ & $N$/$N$(H$_2$O)\,$^b$\\
&\,molec\,cm$^{-2}$& &\\
\hline
(CO$_2$)$_{\rm tot}$ & 26\,$^{c}$ &&&0.335\\
CO$_2$ (in CO) & 1.0\,$^d$ & 0.04 &&0.013  \\
&&&&\\
(CO)$_{\rm tot}$& 16\,$^{c}$ &&&0.195\\
CO (in CO$_2$) & 1.7\,$^e$ &  &0.11&0.021  \\
CO (bc)  & 0.8\,$^f$& &0.05 &0.010  \\
\hline
\end{tabular}
\begin{itemize}
\item[]\footnotesize{Maximum amounts in CO:CO$_2$ mixtures consistent with data. $^{a}$ $A_{\rm \nu_2(CO_2)}$ = $A_{\rm CO}$ = 1.1$\times$10$^{-17}$\,cm\,molec$^{-1}$ \citep{Gerakines1995}. $^{b}$ $N$(H$_2$O) = 8.0$\times$10$^{18}$\,cm$^{-2}$ \citep{Boogert2004}. $^{c}$ From \citet{Boogert2004}. $^d$ Integrated area of CO$_2$ bending mode spectrum at 40\,K in 2:1 CO:CO$_2$ is 1.10\,cm$^{-1}$. $^e$ $\tau_{4.67\,\mu m}$\,=\,0.27\,$\pm$\,0.01, FWHM of 7\,cm$^{-1}$. $^f$ $\tau_{\rm bc}$\,=\,0.30\,$\pm$\,0.01, FWHM of 3.0\,cm$^{-1}$.}
\end{itemize}
\label{fithh46}
\end{table*}
The power of combining CO- and CO$_2$-observational data to elucidate the local environment of both molecules in interstellar ices is illustrated here for the low-mass object HH\,46. Its $\nu_2$(CO$_2$) and CO-stretching mode were recently studied in detail by \citet{Boogert2004}. Here, these bands are re-analysed to establish the maximum possible contribution from a CO-CO$_2$ ice component. \\
\indent Fig.~\ref{hh46}(I,left) shows the $\nu_2$(CO$_2$) spectrum observed with Spitzer \citep[solid black line; CO$_2$ stretching mode not observed,][]{Boogert2004}. The substructure at $\sim$658\,cm$^{-1}$ (15.2\,$\mu$m) shows a weak doublet profile, but the peaks of the doublet are not narrow enough to originate from the $\nu_2$(CO$_2$) band of pure CO$_2$ ice ($T\geq$10\,K). Instead, the more broad profile of the $\nu_2$(CO$_2$) laboratory spectrum that arises when CO$_2$ is mixed in CO at elevated temperatures provides a much better match. Therefore, in Fig.~\ref{hh46}(I) the $\nu_2$(CO$_2$) of HH\,46 is matched by combining the laboratory spectrum of $\nu_2$(CO$_2$) of 2:1 CO:CO$_2$ ice at 40\,K (dotted line) and one at 10\,K in a 10:1 H$_2$O:CO$_2$ mixed ice \citep[similar to that of][light grey dashed line]{Boogert2004}. Grain shape effects may affect the interstellar profile of the first spectrum, although the ice is no longer a homogeneous mixture (FR06b). Since no optical constants are available, these effects
are not taking into account. The overall match (grey line) reproduces the observed band profile, although the small feature at 662\,cm$^{-1}$ and the red-wing beyond 650\,cm$^{-1}$ are less well matched than by \citet{Boogert2004}. The results are listed in Table~\ref{fithh46}. As can be seen from Table~\ref{fithh46}, only 4\% of all CO$_2$ is present in the CO ice (an abundance of 1.3\% with respect to H$_2$O). The majority of the CO$_2$ (96\%) appears to be mixed with the H$_2$O ice. Obviously, this mix and match procedure can only provide a rough estimate of the (maximum) column density of the CO-CO$_2$ ice component contributing to the CO$_2$-bending mode and is only meant to illustrate how laboratory data of binary ices of CO and CO$_2$ can constrain the interstellar ice environment of CO$_2$. \\
\indent Fig.~\ref{hh46}(II,right) shows the spectrum of the solid CO-stretching mode observed (solid line) by VLT-ISAAC \citet{Pontoppidan2003}. Using the same laboratory spectrum (2:1 CO:CO$_2$ at 40\,K, thick solid line in Fig.~\ref{hh46}(IIa)) to fit the blue wing of the CO ice, adding in a 1:10 CO:H$_2$O ice at 10\,K (dashed line) and a pure CO ice component (the mc of interstellar CO-ice, dotted line), gives an abundance of CO in CO:CO$_2$ of 2.1\% with respect to H$_2$O ($\chi^2$\,=\,1.3; see Table~\ref{hh46}). This is 11\% of the total CO present. The results in Fig.~\ref{hh46}(I) and (IIa) show that the interstellar CO and $\nu_2$(CO$_2$) bands of HH\,46 can be fit consistently with spectra of a 
single 2:1 CO:CO$_2$ laboratory ice mixture. In comparison the purely phenomenological fit of the CO band in Fig.~\ref{hh46}(IIb) ($\chi^2$\,=\,1.2) gives a bc (representing the CO:CO$_2$ ice or possibly the LO-TO splitting of pure CO ice; see also Sect.~\ref{discussion}), which is a smaller fraction of the total CO abundance, only 5\% (see Table~\ref{hh46}). 
Further analysis of a much larger sample of sources is needed to distinguish
between these cases and find systematic trends with other parameters such
as temperature. Such systematic studies can then 
address the different formation scenarios of CO$_2$ in
the presence of CO and/or H$_2$O described in Sect.~\ref{introduction},
in particular  whether the amount of CO$_2$ formed with CO is
generally only a small fraction of the total CO$_2$, as found for HH\,46.
\section{Conclusion}
In summary, the experimental work presented here shows that CO only
affects the CO$_2$-bending mode in intimate mixtures below 50\,K
(under the present experimental conditions), where mixing of CO and
CO$_2$ results in a single asymmetric band profile for the CO$_2$
bending mode. In all other CO-CO$_2$ ice morphologies studied here the
CO$_2$-bending mode shows the same doublet profile. Conversely, the
CO-stretching vibration is blue-shifted to a maximum of
2142\,cm$^{-1}$ in intimate mixtures with CO$_2$ and between 2140 and
2146\,cm$^{-1}$ when CO interacts with a layer of initially pure
CO$_2$. The assignment of an interstellar `2143\,cm$^{-1}$ feature' by
CO in layered ices with CO$_2$, however, is difficult. Further
constraints on its assignment require (at least) the analysis of the
interstellar CO$_2$ bending- and future observations of CO$_2$
stretching mode spectra. The laboratory data do indicate that the
combined band profiles of CO and CO$_2$ can be used to distinguish
between mixed, layered and thermally annealed CO-CO$_2$
ices. Ultimately, this can provide important constraints on the
formation mechanism of CO$_2$, one of the most abundant interstellar
ices.

\begin{acknowledgements}
This research was financially supported by the Netherlands Research
School for Astronomy (NOVA) and a NWO Spinoza grant. The authors would
like to thank Klaus Pontoppidan for helpful discussions and Adwin
Boogert for kindly providing the Spitzer CO$_2$ data of HH\,46.
\end{acknowledgements}

\bibliography{bib3}
\bibliographystyle{aa}


\clearpage
\appendix
\section{CO$_2$ asymmetric stretching mode}
\begin{figure*}
\centering
\includegraphics[width=18cm]{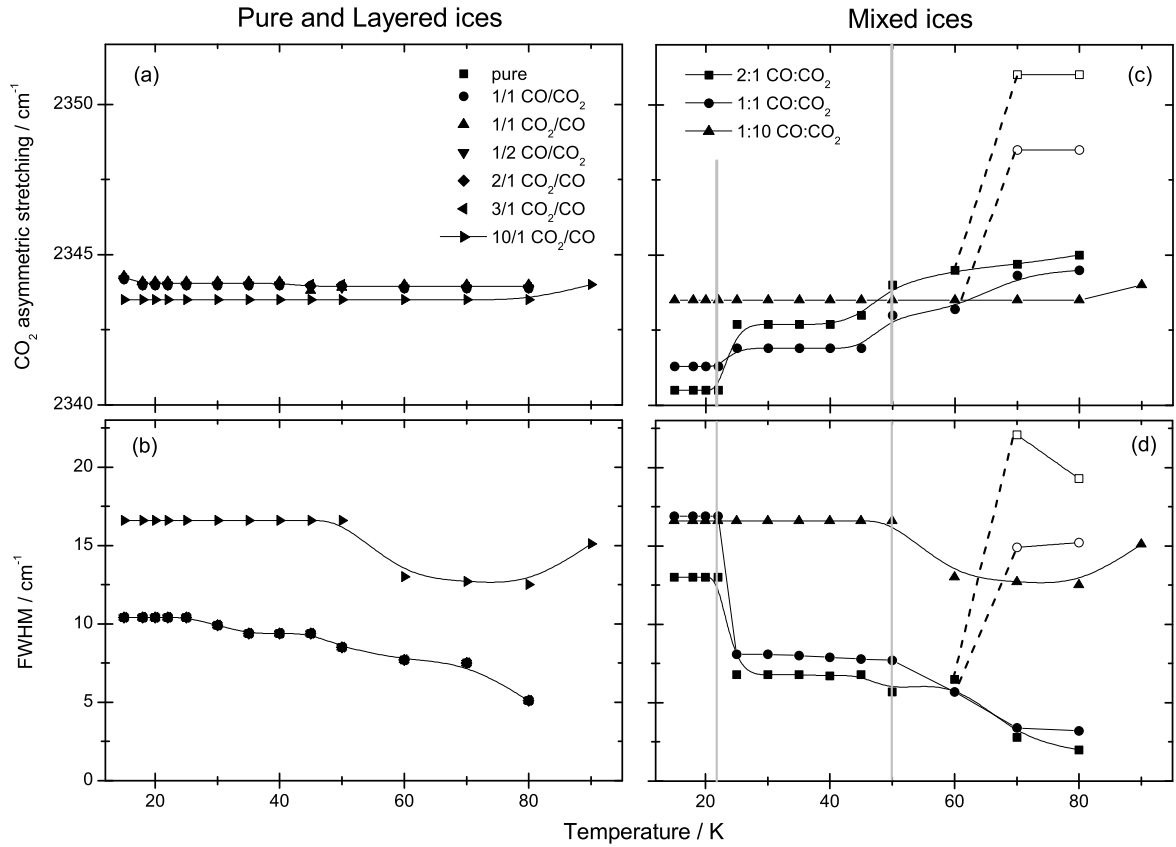}
\caption{The peak-centre positions and FWHM of the main spectral
features of the CO$_2$ asymmetric stretching vibration, plotted as a function of the ice temperature. {\bf (a)} Peak-centre positions in pure and layered
ice with {\bf(b)} the corresponding FWHM. Note that most of the data points overlap in {\bf (a)} and {\bf (b)}. {\bf (c)} Peak-centre positions in mixed ices with {\bf (d)} the corresponding FWHM. The legend in {\bf (a)} gives the symbol assignment for the two left-hand panels, that in {\bf (c)} for the right-hand panels. Open and closed symbols are used to indicate independent peaks present at the same temperature. In each of the panels, spline-fits through the data points guide the eye. In {\bf (c)} and {\bf (d)}, the two vertical grey lines enclose the temperature range between 22 and 50\,K and the dashed lines connect two separate features that appear in the same spectrum. } 
\label{co2stretchpw}
\end{figure*}

\indent In analogy to Figs.~\ref{co2pw} and \ref{copw}, the peak
position and FWHM of all ices studied are summarised in
Fig.~\ref{co2stretchpw} for the CO$_2$ asymmetric stretching mode. 
Sect.~\ref{co2s} describes in detail that the spectral
profile of the CO$_2$ asymmetric stretching mode and its temperature
dependence are very similar for pure and layered ices with CO, but are
significantly different compared to CO:CO$_2$ mixtures. This is even
more clear from Fig.~\ref{co2stretchpw} if the right and
the left hand panels are compared.\\

\indent From Fig.~\ref{co2stretchpw}a, it is clearly seen that
the peak position of
the CO$_2$ asymmetric stretching mode does not change with increasing
temperature in pure or layered ices, although its FWHM
(Fig.~\ref{co2stretchpw}b) does show a small decrease. However, in
mixtures the peak position shows significant shifts to higher
wavenumbers and the evolution of new components as the temperature
increases. This behavioural difference between mixed and layered ices
with CO indicates that (future) observations of the CO$_2$ asymetric
stretching mode in interstellar ice spectra can provide important
additional information on the interstellar ice composition and its
evolutionary history. Because the CO$_2$ asymmetric stretching band is
sensitive to the CO:CO$_2$ environment, spectral information of this
band might also contribute to a better understanding of the nature of
the bc component of interstellar CO-ice (see Sect.~\ref{discussion}).


\end{document}